\documentclass[aps,prl,twocolumn,superscriptaddress,msmath,amssymb]{revtex4}
\usepackage{graphicx}% Include figure files
\usepackage{dcolumn}% Align table columns on decimal point
\usepackage{bm}% bold math

\begin{document}

\title{Spin-valve effect and magnetoresistivity in single crystalline Ca$_3$Ru$_2$O$_7$}

\author{Wei Bao}
%\email{wbao@lanl.gov}
\affiliation{Los Alamos National Laboratory, Los Alamos, New Mexico 87545}
\author{Z.Q. Mao}
\author{Z. Qu}
\affiliation{Department of Physics, Tulane University, New Orleans, Louisiana 70118}
\author{J. W. Lynn}
\affiliation{NIST Center for Neutron Research, National Institute of Standards 
and Technology, Gaithersburg, Maryland 20899}

\date{\today}

\begin{abstract}
The laminar perovskite Ca$_3$Ru$_2$O$_7$ naturally forms ferromagnetic double-layers of alternating moment directions, as in the spin-valve superlattices. The mechanism of huge magnetoresistive effect in the material has been controversial due to a lack of clear understanding of various magnetic phases and phase-transitions.
In this neutron diffraction study in a magnetic field, we identify four different magnetic phases in Ca$_3$Ru$_2$O$_7$ and 
determine all first-order and second-order phase transitions between them. 
The spin-valve mechanism then readily explains the dominant magnetoresistive effect in Ca$_3$Ru$_2$O$_7$.

\end{abstract}


\maketitle

The giant magnetoresistive effect (GMR) of artificial superlattices of magnetic metals\cite{gmr_88,gmr89} has ushered in a new era of spin-based electronics, or spintronics\cite{rev_wolf_01}. It has inspired research on bulk oxides of spin-polarized bands, and led to the discovery of 
the colossal magnetoresistive effect (CMR) 
in perovskite manganites\cite{cmr_93b,cmr_93a,cmr_94}.
The GMR is understood in terms of the spin-valve effect: the resistance
decreases from antiparallel to parallel alignment of neighboring ferromagnetic layers. The CMR in the neighborhood of a ferromagnetic transition is related to the field-tuning of the concomitant ferromagnetic and metal-nonmetal transition, and the many-body double-exchange mechanism is involved\cite{de_zener}.
At low temperatures, smaller but still huge
magnetoresistive effects have been observed in some manganites and can be understood as the spin-valve
effect between ferromagnetic domains\cite{bao94c} or between ferromagnetic layers of the antiferromagnets\cite{msv_98,msv_99}.

Huge magnetoresistive effect has also been observed recently in
the double-layered perovskite Ca$_3$Ru$_2$O$_7$\cite{Cao_Ca327_05l,Cao_Ca327_03B}, which orders antiferromagnetically at $T_N\sim 56$ K\cite{Cao_Ca327_97l} and is isostructural to the manganites investigated in Ref.~\cite{msv_98}. Its behavior at low temperatures is similar to that in the manganites when a magnetic field is applied perpendicular to the easy-axis of magnetization.
A smaller magnetoresistive effect occurs in a first-order metamagnetic transition when the magnetic
field is applied along the easy-axis. The spin-valve mechanism has been proposed in both experimental and theoretical studies\cite{Cao_Ca327_03B,Singh_Ca327_06}. However, the situation is complicated by
a first-order metal-insulator transition at $T_{MIT}\approx 46$ K. In particular, the spin-valve explanation
runs into severe difficulty with the astounding recent experimental conclusion that the fully spin-polarized phase is the least favorable for electronic conduction\cite{Cao_Ca327_05l}. Thus, alternative mechanisms such as orbital order have been proposed\cite{Cao_Ca327_05l,Cooper_Ca327_04}.

The field-induced ``ferromagnetic'' phase, which should be called the paramagnetic phase\cite{def}, as well as the antiferromagnetic phases of 
Ca$_3$Ru$_2$O$_7$, had been deduced from bulk measurements\cite{Cao_Ca327_03A}. They
form the foundation in current understanding of the novel 
magnetoresistive phenomena in Ca$_3$Ru$_2$O$_7$\cite{Cao_Ca327_03B,Singh_Ca327_06,Cao_Ca327_05l}. However, magnetic structures in these phases have not been directly determined using microscopic techniques. Here we report a magnetic neutron diffraction study of all phases of Ca$_3$Ru$_2$O$_7$ in the temperature $T$ and magnetic field $B$ plane. With the magnetic structures and phases correctly identified, our results offer direct support for the spin-valve mechanism as the origin of the huge magnetoresistive effect in Ca$_3$Ru$_2$O$_7$.

Neutron diffraction experiments were performed at NIST using the thermal neutron triple-axis spectrometer BT9 in the triple-axis mode. Both monochromator and analyzer used the
(002) reflection of pyrolytic graphite to select neutrons of energy $E=14.7$ meV. Pyrolytic graphite filters of 10 cm total thickness were
placed in the neutron beam path to reduce higher order neutrons.
The horizontal collimations were 40$^{\prime}$-48$^{\prime}$-44$^{\prime}$-100$^{\prime}$. A single-crystal sample of Ca$_3$Ru$_2$O$_7$ of 0.73 g was used. The plate-like sample was cleaved from a larger piece grown using a floating-zone method in a mirror-focused furnace. The lattice parameters of the
orthorhombic unit cell (space group No.~36 $Bb2_1m$\cite{Cao_Ca327_00B,Yoshida_Ca327_PRB}) are $a=5.366$, $b=5.522$, and $c=19.49 \AA$ at 
4.5 K. There are four Ca$_3$Ru$_2$O$_7$ per unit cell. 
The intergrowth of other members of the Ruddlesden-Popper series, which have similar lattice parameters $a$ and $b$ but very different $c$, does not occur in our sample. The single crystal sample was first aligned in the ($0kl$) zone for measurements and
the sample temperature was regulated by a top-loading pumped He cryostat.
The same sample was re-aligned in the ($h0l$) zone and the temperature and magnetic field were regulated by a 7 T vertical field cryomagnet. All magnetic phases of Ca$_3$Ru$_2$O$_7$ can be
accessed in these experimental configurations\cite{corr}.

\begin{figure}[tbh]
\includegraphics[width=0.6\columnwidth,angle=90,clip=]{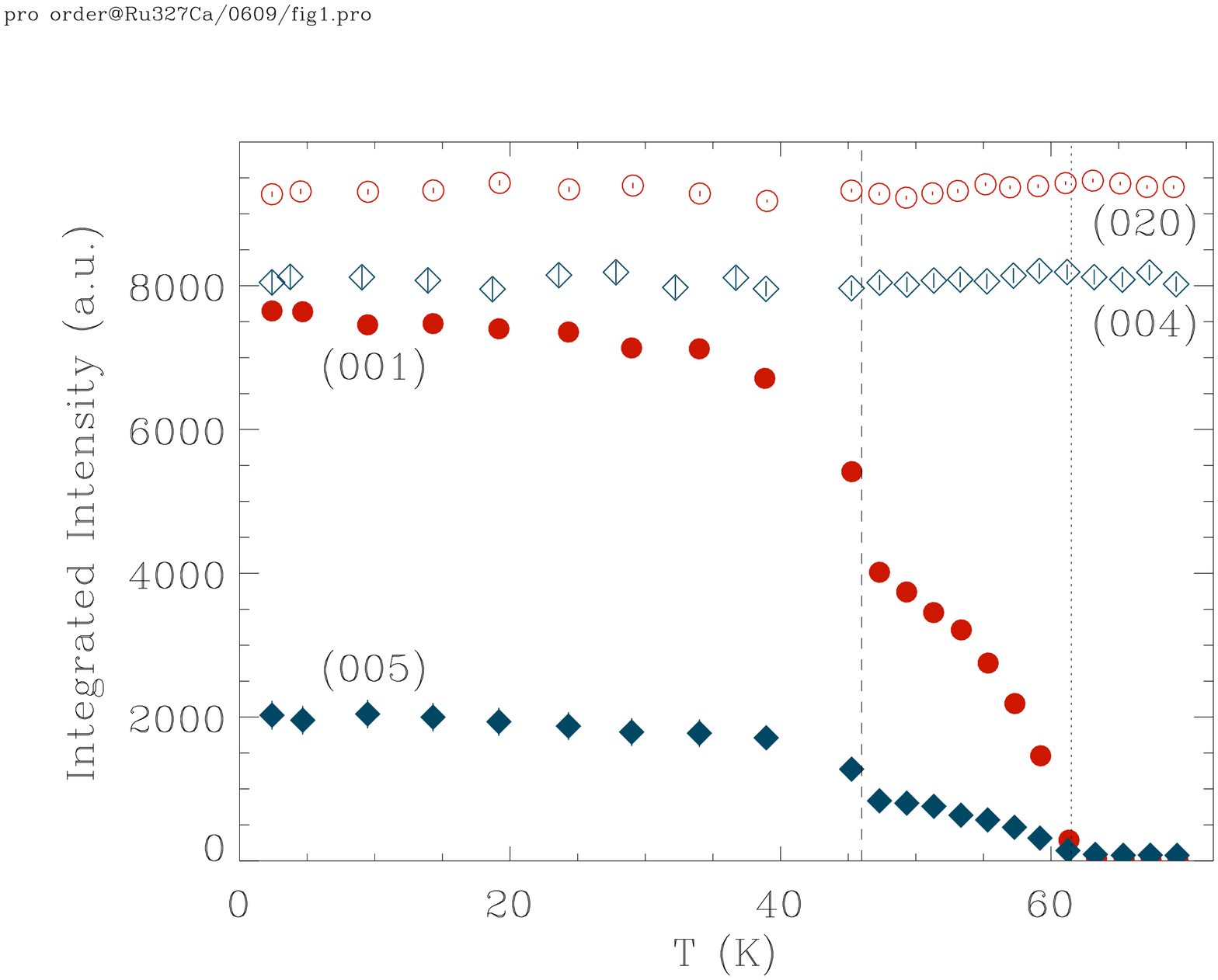}
\vskip -5mm
\caption{Temperature dependence of the magnetic Bragg intensity at (001) and (005) and structural Bragg intensity at (004) and (020). The dashed line denotes the metal-insulator transition at $T_{MIT}=46$ K, and the dotted line the antiferromagnetic transition at $T_N=62$ K.}\label{fig1}
\end{figure}

Fig.~\ref{fig1} shows the integrated intensity of rocking scans at selected structural and magnetic Bragg peaks as a function of rising temperature at $B=0$.
The antiferromagnetic peaks appear below $T_N=62$ K and get enhanced by the metal-to-insulator transition at $T_{MIT}=46$ K. Meanwhile, there is no clear anomaly at either $T_N$ or $T_{MIT}$ for the structural Bragg intensities, corroborating the conclusion of a previous powder neutron 
diffraction study that the space-group symmetry of the crystal is not affected by the phase transitions, although the lattice parameters experience a first-order-like change at $T_{MIT}$\cite{Yoshida_Ca327_PRB}.

\begin{figure}[b]
\includegraphics[width=0.7\columnwidth,angle=90,clip=]{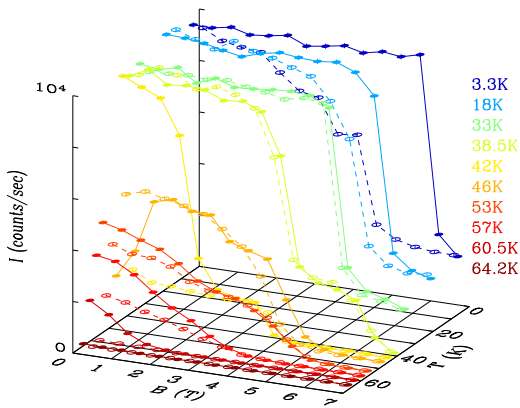}
\vskip -5mm
\caption{Antiferromagnetic (001) Bragg peak as a function of magnetic field along the $b$-axis at some representative temperatures. The solid circles were measured with the field increasing, and the open ones with the field decreasing.}\label{fig2}
\end{figure}

Fig.~\ref{fig2} shows the antiferromagnetic (001) peak as a function of magnetic field applied along the $b$-axis at some representative temperatures.
The sample was cooled in zero field. The Bragg intensity was measured first with the field ramping up, then back down to zero. The sample temperature was then raised to the next measurement point. Below $T_{MIT}$, corresponding to the metamagnetic transition, the antiferromagnetic intensity drops at the first-order transition. The high
field phase was previously regarded as the completely spin-polarized ``ferromagnetic'' state\cite{Cao_Ca327_05l,Cao_Ca327_03B,Cao_Ca327_00B,Cao_Ca327_03A,Cooper_Ca327_04}. However, about 20\% of the (001)
intensity in our measurements survives the metamagnetic transition. It disappears only
in a second-order transition at higher temperatures and fields. For $T \gtrsim T_{MIT}$ where the  metamagnetic transition no longer
occurs, the antiferromagnetic Bragg peak shows a different type of hysteresis behavior at low
magnetic field. The field-sweeping measurements (Fig.~\ref{fig2}) are corroborated by temperature-sweeping measurements as in Fig.~\ref{fig5}(d).

\begin{figure}[tb]
\center{
\includegraphics[width=0.6\columnwidth,angle=90,clip=1]{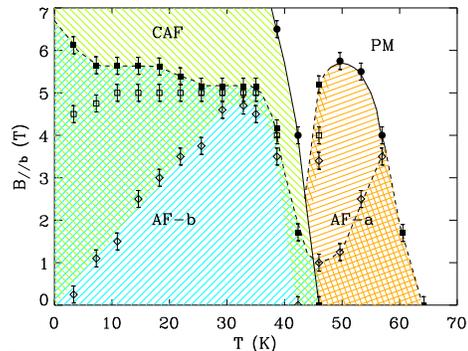}
}
\vskip -7mm
\caption{ Phase diagram of Ca$_3$Ru$_2$O$_7$ in a magnetic field applied along the $b$-axis. The solid lines denote continuous phase-transitions, and dashed lines first-order ones. The hysteresis regions are marked by cross-lines. Magnetic structures in the four phases are shown in Fig.~\ref{fig4}.}\label{fig3}
\end{figure}

The phase diagram resulting from our measurements is shown in Fig.~\ref{fig3}. The continuous phase transitions are delimited by the circles and solid lines. The solid squares and broken lines delimit
the first-order phase transitions, and the hysteretic regions are marked by open diamonds and are cross-hatched. 
The peak positions of $dI/dB$ in the field-down measurements (Fig.~\ref{fig2}) are also provided in Fig.~\ref{fig3} by the open squares. There exists three antiferromagnetic phases, labeled $AF$-$a$, $AF$-$b$ and $CAF$ in the shaded regions of the phase diagram, which magnetic structures we determined.

In the previous powder neutron diffraction work at zero field, only one magnetic Bragg peak at (001) can been observed at low temperatures\cite{Yoshida_Ca327_PRB}. Although it determines the antiferromagnetic wavevector in the $AF$-$b$ phase, the complete magnetic structure such as the direction and size of the magnetic moment was not determined. 
We have measured 15 independent magnetic Bragg peaks at 3.5~K and $B=0$ in this single crystal. The integrated intensity is normalized to absolute units using structural Bragg peaks. The resulting
magnetic cross-section $\sigma_{obs}$ 
is presented in Table I. 
\begin{table}[b]
\caption{Magnetic Bragg cross-section, $\sigma_{obs}$,
observed at 3.5~K and $B=0$ in units of barns per unit cell.
The theoretical cross-section $\sigma_{I}$
in the same units is calculated for the ``$++--$''
magnetic structure, and $\sigma_{II}$ for the ``$+--+$''
magnetic structure, using $M=1.8(2)$ $\mu_B$/Ru pointing along the b-axis. The former is supported by our measurements.
}
\label{mlist}
\begin{ruledtabular}
\begin{tabular}{clccc}
${\bf q}$ & $\sigma_{obs}$ & $\sigma_{I}$ &  $\sigma_{II}$  \\
\hline
      (0      0      1) &      9.64(3)   &   9.7932   &   5.0210 \\
      (0      0      3) &      1.10(2)  &   0.9802   &   10.737\\
      (0      0      5) &      5.91(6)  &    7.4051  & 0.0089\\
      (0      0      7) &     0.43(2)   &  0.4805   &   3.3394\\
      (0      0      9) &      1.34(5)  &   0.9638  &   0.6579\\
      (0      0      11) &     0.63(4) &    0.4037  &   0.1533\\
      (0      2      1) &    0.070(7)  &  0.0473  &  0.0242\\
      (0      2      3) &    0.051(7)  &  0.0375  &   0.4106\\
      (0      2      5) &     0.44(2)  &   0.6375 & 0.0008\\
      (0      2      7) &    0.05(1)  &  0.0630   &  0.4379\\
      (0      2      9) &     0.39(3)   &  0.1565  &  0.1069\\
      (0      2      11) &     0.27(3)  &  0.0679  &  0.0258\\
      (2      0      1) &      2.39(1)  &    2.2154  &    1.1358\\
      (2      0      3) &     0.216(5)  &   0.2261   &   2.4762\\
      (2      0      5) &      1.66(1)  &    1.7618 &  0.0021\\
\end{tabular}
\end{ruledtabular}
\end{table}
The selection rule for the magnetic Bragg peaks indicates that the magnetic moments in each RuO$_2$ plane are ferromagnetically aligned, and the odd $l$ index restricts possible
collinear magnetic structure to two choices:
I) The RuO$_2$ bilayers are ferromagnetically aligned, and the bilayers are antiferromagnetically stacked. The sign of the magnetic moment thus follows the $++--$ sequence in consecutive layers in the unit cell 
(see the $AF$-$a$ or $AF$-$b$ in Fig.~4).
\begin{figure}[tb]
\includegraphics[width=0.7\columnwidth,angle=-90,clip=]{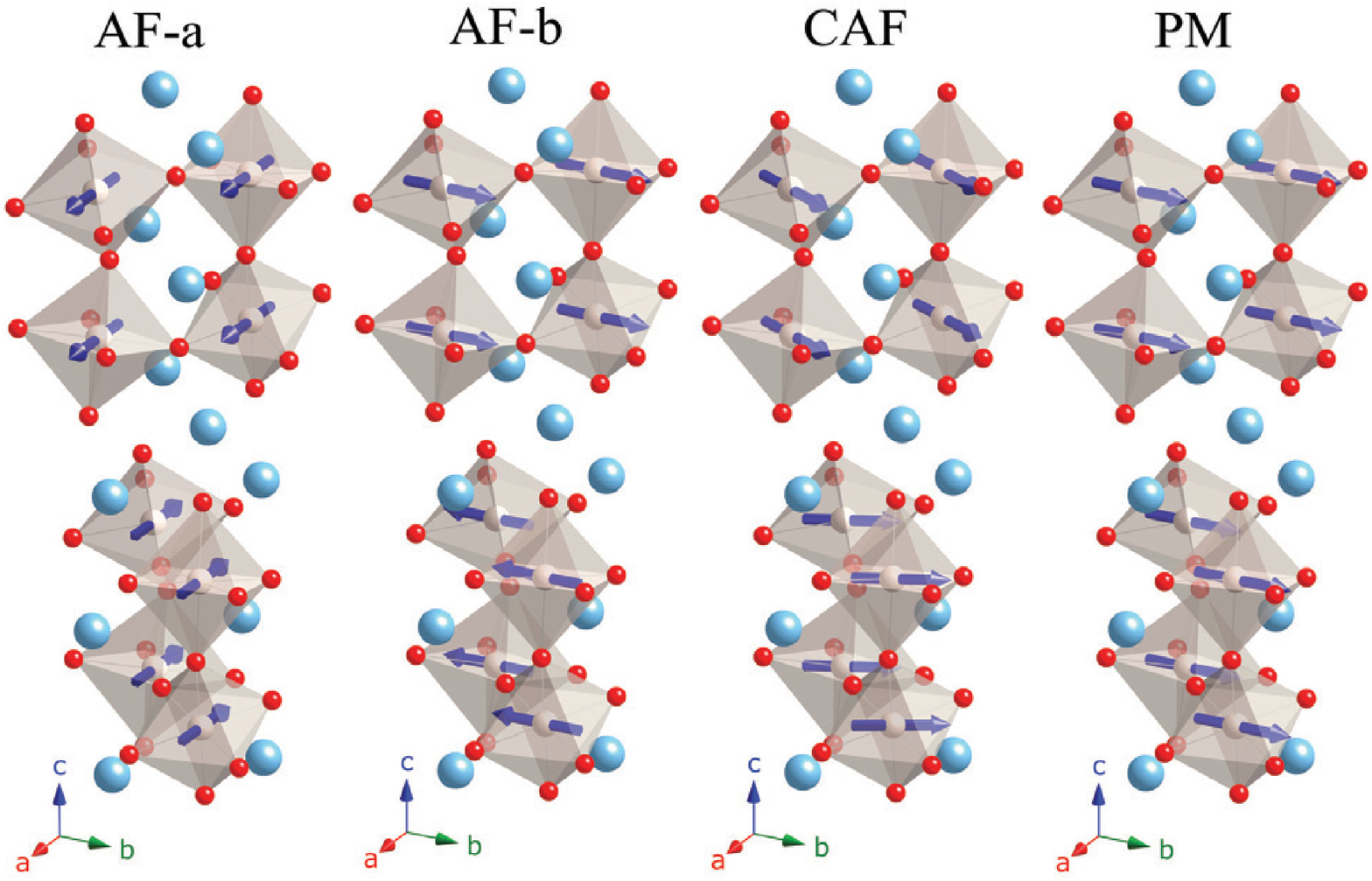}
\vskip -3mm
\caption{Magnetic structures in the $AF$-$a$, $AF$-$b$ and $CAF$ phases. In the paramagnetic phase (PM), all magnetic moments are aligned by the applied magnetic field.}\label{fig4}
\end{figure}
II) The antiferromagnetically aligned bilayers stack antiferromagnetically. Thus the sign follows the $+--+$ sequence. The antiferromagnetic structure which follows the $+-+-$ stacking sequence is forbidden by the selection rule.

The magnetic neutron diffraction cross-section is given by
\begin{equation}
\sigma({\bf q})=\left(\frac{\gamma r_0}{2}\right)^2
	\langle M\rangle^2 \left|f(q)\right|^2 
	[1-(\widehat{\bf q}\cdot \widehat{\bm{M}})^2]
	\left| {\cal F}({\bf q})\right|^2,
\label{eq_cs}
\end{equation}
where $(\gamma r_0/2)^2=0.07265$~barns/$\mu_B^2$, $\bm{M}$ is the
magnetic moment of the Ru$^{4+}$ ion, $f(q)$ the Ru$^{4+}$ magnetic
form factor, $\widehat{\bf q}$ the unit vector of ${\bf q}$, and $\left| {\cal F}({\bf q})\right|^2$ the squared magnetic structure factor\cite{neut_squire}. $\left| {\cal F}({\bf q})\right|^2$ depends mainly on the $l$ index with
\begin{equation}
\left| {\cal F}(20l)\right|^2 \approx \left| {\cal F}(02l)\right|^2 \approx \left| {\cal F}(00l)\right|^2 =
	64 \cos^2(l\epsilon)
\label{eq_I}
\end{equation}
for model I, and 
\begin{equation}
\left| {\cal F}(20l)\right|^2 \approx \left| {\cal F}(02l)\right|^2 \approx \left| {\cal F}(00l)\right|^2 =
	64 \sin^2(l\epsilon)
\label{eq_II}
\end{equation}
for model II, where $2\epsilon=0.1978 c$ is the bilayer separation\cite{Yoshida_Ca327_PRB}. The substantial difference
between the observed $\sigma_{obs}$ of (20$l$) and (02$l$) in Table I thus is due to
$ \bm{M}$ being parallel to $\bm{b}$. The $l$ dependence of $\left| {\cal F}({\bf q})\right|^2$ has opposite phases for the two different models in Eq.~(\ref{eq_I})-(\ref{eq_II}), and the observed pattern in Table I supports model I. A least-squared fit of
measured cross-sections to Eq.~(\ref{eq_cs})-(\ref{eq_I}) yields a magnetic moment $M=1.8(2)\mu_B$ per Ru with the $b$-axis as the easy-axis\cite{corr}. The magnetic structure for the $AF$-$b$ phase is depicted in Fig.~\ref{fig4}.

The metal-insulator transition at $T_{MIT}$ drastically affects the magnetic Bragg intensities
(Fig.~\ref{fig1}). However, the fractional reduction of the intensities from 3.5 to 50.5 K at magnetic Bragg peaks (00$l$), $l$=1,3,\dots,11 is the same. Therefore, the stacking sequence of antiferromagnetic moments above $T_{MIT}$ in the $AF$-$a$ phase remains the same. On the other hand, the relative intensities of the (201) and (021) Bragg peaks switch at
$T_{MIT}$, see Fig.~\ref{fig5}(a) and (b).
\begin{figure}[tb]
\includegraphics[width=0.65\columnwidth,angle=90,clip=]{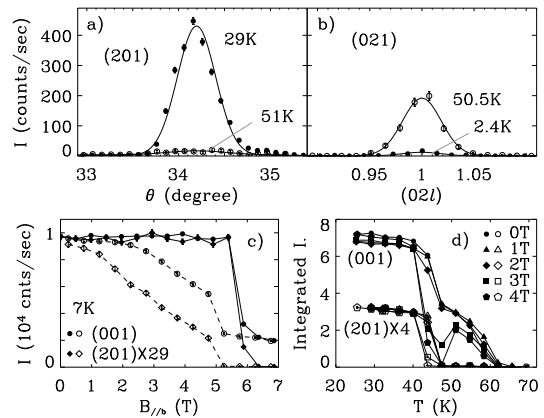}
\vskip -3mm
\caption{(a)-(b) The (201) and (021) magnetic Bragg peaks, which are only $\sim$8$^{o}$ away from the $a$ and $b$-axis, respectively, below and above $T_{MIT}=46$ K at $B$=0. 
(c) Field dependence of the (001) and (201) Bragg intensities at 7 K. 
(d) Temperature dependence of the (001) (solid symbols) and (201) (open symbols) Bragg intensities at various field strengths. }\label{fig5}
\end{figure}
Thus, the easy axis changes from the $b$-axis to the $a$-axis at the metal-insulator transition, in excellent agreement with bulk data analysis\cite{Cao_Ca327_03A}. In both the $AF$-$a$ and $AF$-$b$ phases, $\widehat{\bf q}\cdot \widehat{\bm{M}}=0$ in Eq.~(\ref{eq_cs}) for the (00$l$) antiferromagnetic peaks. These peaks, as in Fig.~\ref{fig1}, hence represent the squared order parameter $\langle M(T)\rangle^2$ of the magnetic transitions, irrespective of the change of the easy axis in the basal plane.

At low temperatures in the $AF$-$b$ phase, the magnetic field applied along the easy $b$-axis induces
a metamagnetic transition near 6 T \cite{Cao_Ca327_03A,Cao_Ca327_05l}. The ferromagnetic moment
is 1.73$\mu_B$/Ru at 5 K, very close to the 2$\mu_B$ saturation moment of the Ru$^{4+}$
($4d^4$) low-spin $S$=1 ion\cite{Cao_Ca327_03B}. However, the high-field $CAF$ state is not a fully polarized ferromagnetic state
as previously presumed\cite{Cao_Ca327_05l,Cao_Ca327_03B,Cao_Ca327_00B,Cao_Ca327_03A,Cooper_Ca327_04}, since substantial intensity of the (001) antiferromagnetic peak remains, see Fig.~\ref{fig2}. In Fig.~\ref{fig5}(c), the field dependence
of the (001) and (201) peaks at 7 K, during the field ramping up (solid symbols) and down (open symbols), are compared. The absence of the (201) peak in the $CAF$ phase indicates that the reduced antiferromagnetic component,
$1.8(2)\times [I(7{\rm T})/I(0{\rm T})]^{1/2}=0.80(9) \mu_B$/Ru from the (001) peak, has switched its direction from the $b$-axis to the 
$a$-axis at the metamagnetic transition. Thus the magnetic structure in the $CAF$ phase is a canted antiferromagnet: a vector sum of an $AF$-$a$ antiferromagnet with a staggered moment of 0.80 $\mu_B$
and a ferromagnet with a moment of 1.73 $\mu_B$ along the $b$-axis, resulting in a canting angle 25$^o$ from the $b$-axis (see Fig.~\ref{fig4}). The total staggered magnetic moment is 1.91(4)$\mu_B$/Ru, which is within the error bar of the staggered moment at the same temperature in the $AF$-$b$ phase.

With the phase diagram and the magnetic structures of Ca$_3$Ru$_2$O$_7$ determined, bulk magnetic data\cite{Cao_Ca327_05l,Cao_Ca327_03B,Cao_Ca327_00B} can be readily understood. Below $T_{MIT}$, the magnetic field applied along the easy $b$-axis would induce
a first-order spin-flop transition. With increasing magnetic field, there is a second order phase transition to the paramagnetic phase as the field-induced canting angle decreases continuously to zero. When the external field is applied perpendicular to the $b$-axis, there would be only a second-order phase transition to the paramagnetic phase as the antiferromagnetic moments continuously rotate towards the field direction. Such behavior is what one would expect from an antiferromagnet with modest magnetic anisotropy, as has been observed in MnF$_2$\cite{MnF2} and GdAlO$_3$\cite{GdAlO3}. Above $T_{MIT}$, the metal-insulator transition alters the moment orientation of Ca$_3$Ru$_2$O$_7$ to the $a$-axis. Now the first-order metamagnetic spin-flop transition is expected when the field is applied along the $a$-axis, while a second-order transition is expected
when the field is along the $b$-axis. The latter is only partially fulfilled in Ca$_3$Ru$_2$O$_7$ as the single-phased $AF$-$a$
region is unusually bounded by mixed phase regions on both the low and high temperature sides (see Fig.~\ref{fig3}). The hysteresis is originated from the first-order metal-insulator transition, and can be a source of confusion in interpreting bulk data if the thermal and field history are not carefully monitored.

The magnetic structures in the $AF$-$a$ and $AF$-$b$ phases differ only in their easy axis. The switching of moment direction is
associated with the metal-insulator transition. This association offers the
opportunity of tuning transport with magnetic field by driving the system across the phase boundary between the $AF$-$b$ and $AF$-$a$ phases in the vicinity of $T_{MIT}$ (Fig.~3), like the CMR in manganites.
Hence, the $c$-axis magnetoresistivity $\rho_c(T,B)$ in the neighborhood of $T_{MIT}$\cite{Cao_Ca327_05l} should be understood in terms of the metal-insulator transition. On the other hand, the huge 
negative magnetoresistivity away from the metal-insulator transition can be readily understood by the spin-valve effect: the closer to a ferromagnetic moment arrangement, the lower the resistivity. The spins are not fully polarized by the external field along the $b$-axis in the $CAF$ phase, while they are in the paramagnetic phase. In addition to the mis-characterization of the $CAF$ phase, the downward extension of the paramagnetic phase near $T_{MIT}$ and the hysteresis caused by the first-order metal-insulator transition also have been sources of confusion.

In summary, the first-order metal-insulator transition considerably complicates magnetic behavior of Ca$_3$Ru$_2$O$_7$, and has made interpreting bulk magnetic data difficult. Using neutron diffraction, we have successfully determined the phase-diagram and magnetic structures. The huge magnetoresistive effect in Ca$_3$Ru$_2$O$_7$ now can be readily understood by the spin-valve effect. At the metal-insulator transition, magnetoresistive effect is further effected by the
concomitant magnetic transition between different magnetic phases.

Work at LANL was supported by the US DOE, at Tulane 
by NSF under grant DMR-0645305, by DOE under DE-FG02-07ER46358
and by the Research Corp.


\begin{thebibliography}{23}
\bibitem{corr}
In denoting the in-plane direction of applied magnetic field, the $a$ and $b$ axes in \cite{Cao_Ca327_05l,Cao_Ca327_03B,Cao_Ca327_00B,Cao_Ca327_03A,Cooper_Ca327_04} should switch in order to be consistent with
the lattice-parameter definition therein.
\end{thebibliography}
\end{document}